\documentclass[prl,twocolumn,showpacs]{revtex4}
\usepackage{epsfig}
\usepackage{graphicx}
\usepackage{color}

\begin{document}

\newcommand{\im}{{\rm Im}}
\newcommand{\re}{{\rm Re}}
\newcommand{\up}{\uparrow}
\newcommand{\dn}{\downarrow}
\newcommand{\eps}{\varepsilon}
\newcommand{\ave}[1]{\langle #1\rangle}

\title{Correlation driven Mott-insulator-to-metal transition}

\author{Rubem Mondaini$^{1,2}$ and Thereza Paiva$^3$}
\affiliation{$^1$Department of Physics, Georgetown University, Washington, District of Columbia 20057, USA}

\affiliation{$^2$Physics Department, The Pennsylvania State University, 104 Davey Laboratory, University Park, Pennsylvania 16802, USA}

\affiliation{$^3$ Instituto de F\'\i sica, Universidade Federal do Rio de Janeiro Cx.P. 68.528, 21941-972 Rio de Janeiro RJ, Brazil}

\begin{abstract}
We study transport properties of the half-filled two-dimensional ($2D$) Hubbard model with spatially varying interactions, where a pattern of interacting and non-interacting sites is formed.  We use Determinantal  Quantum Monte Carlo method to calculate the  double occupation, effective hopping and Drude weight. These data point to two phase transitions, driven by fermionic correlations. The first is the expected  metal to a Mott insulating state. The second one, is an exotic transition from a Mott insulating state to a highly anisotropic metal, that takes place at large values of the fermion-fermion interaction. This second transition occurs when the layers formed by the spatially varying interactions decouple due to the suppression of the hopping between interacting and non-interacting sites, leading to fermionic transport along the non-interacting ones.
\end{abstract}
\maketitle

Metal insulator transitions have been a topic of intense study over many years\cite{nandini}.  When the transition from a metallic into an insulating  state is driven by  correlations
it is known as the Mott transition \cite{Mott,imada}.   The experimental observation of the Mott  insulating state in two-flavor mixtures of fermionic atoms loaded on optical lattices, \cite{Bloch08,Esslinger} opened new possibilities in this field.
Models that take correlation effects in tight-binding systems, such as the Hubbard model, have been widely used to describe different types of insulating phases and metal-insulator transitions.

An interesting point that has been addressed in the literature is whether these same correlations could drive an insulating system metallic.
When a periodic  on-site energy is distributed in the system,  we have the so called Ionic Hubbard Model. In the non-interacting limit, the periodic potential
produces a dispersion relation that is gapped at half-filling.  In these systems,  correlations are responsible for a band-insulator to metal transition.  \cite{ionic-rts,ionic-randeria,ionic-dagotto}.

The same on-site energies can be randomly distributed throughout the lattice, leading to an  Anderson insulator \cite{lee85,belitz94} depending on the strength of the disorder and system dimensionality. The use of speckle noise \cite{aspect,demarco} in cold atoms loaded on optical lattices has enabled a new experimental route to  understanding  the interplay between disorder and interactions.  This is still an area of intense debate, with many open questions. Nonetheless, numerical data for the Anderson-Hubbard model  points to an Anderson insulator to metal transition driven by correlations \cite {anderson-hubbard}.

Can correlations drive a {\it Mott insulator}  into a metal?  Here we show that, when the interactions are not homogeneously distributed in the system, but instead are spatially varying, this correlation induced Mott insulator to metal transition can indeed take place.   Although spatially varying interactions are not yet available on optical lattices, they have recently become available in ultracold gases. The ability to control a magnetic Feshbach resonance with laser light \cite{bauer}, has increased the tunability of interactions for bosonic systems. Submicron spatial modulation of the interaction was already achieved  in a $^{174}$Yb Bose-Einstein condensate \cite{boson-sl}. Optical  control of Feshbach resonances for a fermionic ultracold gas \cite{fermion-sl} has also  been proposed.

\begin{figure}[t]
\vspace{-0.9cm}
{\centering\resizebox*{2.6in}{!}{\includegraphics*{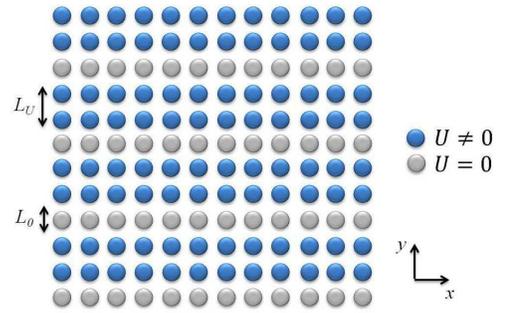}}}
\vspace{-0.3cm}
\caption{(Color online) Schematic figure of the spatially varying interactions in a system  with $L_U=2$ $L_0=1$ in a $12\times12$ lattice.
}
\label{fig:sl} 
\end{figure}

Here we consider a modified version of the Hubbard Model (HM) with spatially varying interactions, whose Hamiltonian reads
\begin{eqnarray}
\mathcal{H} &=& - t \sum_{\langle i,j\rangle,\sigma}
(c^\dagger_{i\sigma} c_{j \sigma}^{\phantom\dagger} +
c^\dagger_{j\sigma} c_{i \sigma}^{\phantom\dagger} )
\nonumber
\\
&&+ \sum_i U_i\left(n_{i\uparrow}-\frac{1}{2}\right)\left(n_{i\downarrow}-\frac{1}{2}\right)
-\mu \sum_{i,\sigma} n_{i\sigma},
\label{eq:hamil}
\end{eqnarray}
where, in standard notation $c^\dagger_{i\sigma} $ ($c_{i\sigma} $) creates (destroys) a fermion in site $i$ in state $\sigma$.
The hopping parameter between nearest neighbor sites $(\langle i,j\rangle)$  is set to $t=1$,  $U_i$ is the site dependent  interaction, and $\mu$ is the chemical potential controlling the band filling to yield a given fermionic density $\rho$. The interaction term is written in particle-hole symmetric form. Thus, tuning $\mu=0$ drives the occupation to one in every site for all Hamiltonian parameters $t$, $U_i$ and temperatures $T$. We have here restricted our study to half-filled systems. To simulate the spatially modulated systems we construct a layered pattern of repulsive and non-interacting layers where $U>0$  and $U=0$, respectively.
We define the width of the repulsive layer as $L_U$ and that of the non-interacting  one as $L_0$ as depicted in Fig. \ref{fig:sl} for a $L_U=2$ $L_0=1$  in a $12\times12$ lattice. Note that not  all  patterns  fit all the available  lattice sizes.
Here we use Determinantal Quantum Monte Carlo (DQMC) simulations \cite{dqmc}  to probe  transport  properties of the half-filled  two-dimensional square $L \times L$ system with spatially varying interactions. Metal-insulator transitions of one-dimensional non-symmetrical Hubbard superlattices \cite{TCLP96} were
studied and shown to have an interesting behavior where the Mott transition takes place away from half-filling, at a pattern dependent density \cite{TCLP98}.

A quantity of interest is the double occupation $ d=\langle n_{\uparrow}n{_\downarrow}\rangle $, as it is measured in optical lattices experiments. Indeed, it has been used to probe metal insulator transitions \cite{Esslinger} in fermionic two-flavor mixtures of $^{40}K$ atoms. For the homogeneous, non-interacting, metallic system $ d=\langle n_{\uparrow}\rangle\langle n{_\downarrow}\rangle =0.25$. As correlations are turned on at a fixed temperature, $d$ decreases, signaling the metal-insulator transition, and approaches 0 as $U \to \infty$.
 The symmetric form of the Hamiltonian requires that, at half-filling, the charge distribution is homogeneous throughout the lattice.  The double occupation, on the other hand, is not homogeneous and follows the  superlattice pattern.  Single-atom resolved double occupation \cite{singleatom} is experimentally available, therefore it is then relevant to define  $d_0$ and $d_U$,  the average double occupation within non-interacting and interacting sites, respectively.

\begin{figure}[t]
\vspace{-1.0cm}
{\centering\resizebox*{3.0in}{!}{\includegraphics*{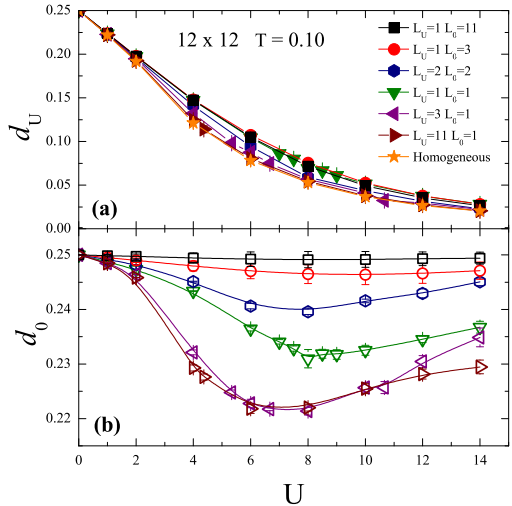}}}
\vspace{-0.7cm}
\caption{(Color online) Double occupation within (a) repulsive ($d_U$) and (b) non-interacting sites ($d_0$) as a function of $U$ for different $12 \times 12$ superlattices  at  $T=0.10$. Double occupation for the corresponding homogeneous system (stars) is also shown in (a) for comparison.}
\label{fig:DoccvsU} 
\end{figure}

The $U$-dependence of both $d_0$ and $d_U$ for different superlattice patterns and fixed  temperature ($T=0.10$) is shown in Fig. \ref{fig:DoccvsU}. Within the  repulsive sites  (Fig. \ref{fig:DoccvsU}(a)) $d_U$ decreases monotonically with $U$, and approaches that of  the  corresponding homogeneous system as the  ratio $L_U/L_0$ is increased. The  behavior of $d_U$ is in agreement with a metal-insulator transition as the interactions are turned on. The double occupation on the non-interacting sites (Fig. \ref{fig:DoccvsU}(b)) is affected by the strength of the interaction on the neighboring repulsive ones and displays a surprising non-monotonic behavior with $U$. Starting from $U=0$, as $U$ increases the effect of the interactions ``leaks" into the non-interacting sites. When the ratio $L_U/L_0$ increases, the double occupation is reduced even where the interaction is turned off.
It is worth noting the difference in the cases $L_U=1$ $L_0=1$ and $L_U=2$ $L_0=2$, where the ratio $L_U/L_0$ is the same, but the wider free layer in the latter makes the spins more uncorrelated in the non-interacting region.
As $U \rightarrow\infty$  double occupations are forbidden in the repulsive sites and  hopping between  free and repulsive sites is completelly suppressed,
leading to a decoupling between free and repulsive layers and bringing $d_0$ back to the value of the homogeneous non-interacting system,
 $d_0 \to 0.25$.
The upturn in $d_0$ signals the onset of the  decoupling between the  layers, that, as we shall see bellow,   leads to an insulator-to-metal transition.

\begin{figure}[t]
\vspace{-1.0cm}
{\centering\resizebox*{3.5in}{!}{\includegraphics*{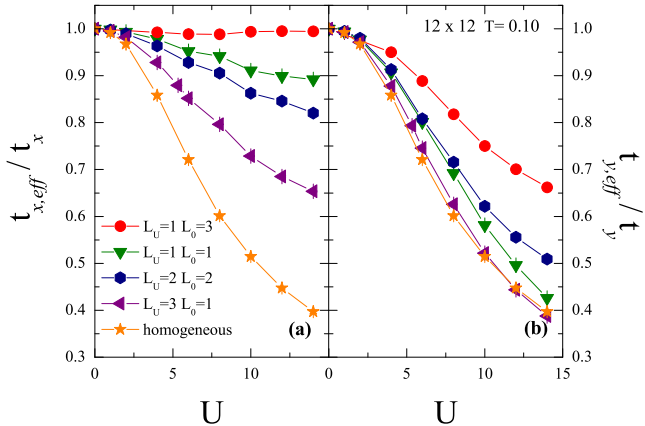}}}
\vspace{-0.5cm}
\caption{(Color online)
Effective hopping along (a) and across (b) the direction of the layers as a function of interaction strength  $U$,  for  $12 \times 12$ superlattices  at  $T= 0.10$. Results for the corresponding homogeneous system (stars) are shown in both panels for comparison.
}
\label{fig:txy_effvsU} 
\end{figure}

We now analyze the
effective hopping \cite{White89b,Varney09},
which is defined as  the ratio of the kinetic energy on a superlattice to its non-interacting homogeneous counterpart value, both  in the direction  along  ($x$) and across ($y$) the layers;
\begin{equation}
\frac {t_{\alpha,eff}} {t_\alpha}=
\frac {\langle c^\dagger_{j+\hat{\alpha}\sigma} c_{j \sigma}^{\phantom\dagger}+ c^\dagger_{j\sigma} c_{j +\hat{\alpha}\sigma}^{\phantom\dagger}  \rangle_{SL}}  {\langle c^\dagger_{j+\hat{\alpha}\sigma} c_{j \sigma}^{\phantom\dagger}+ c^\dagger_{j\sigma} c_{j +\hat{\alpha}\sigma}^{\phantom\dagger}  \rangle_{U=0}},
\end{equation}
where $\alpha=x$ or $y$.
One would expect that anisotropy favors the fermionic transport along the direction of the layers. This is indeed the case, as can be  readily observed in Fig. \ref{fig:txy_effvsU}, where the repulsive interaction splits the $x$ and $y$ contributions to the effective hopping, reducing $t_{y,eff}/t_y$ more strongly  than $t_{x,eff}/t_x$ as $U$ is increased. In the direction perpendicular to the layers the effective hopping approaches that of the corresponding homogeneous system, as $L_U/L_0$ increases, displaying the characteristic behavior of an insulating system. Along the direction of the layers, on the other hand, the effective hopping is close to that of the non-interacting  metallic system when $L_U/L_0 <1$.

\begin{figure}[t]
\vspace{-1.0cm}
{\centering\resizebox*{3.2in}{!}{\includegraphics*{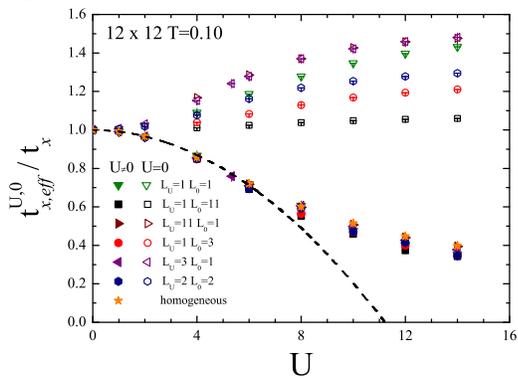}}}
\vspace{-0.5cm}
\caption{(Color online)
Effective hopping along  free (open symbols) and repulsive (closed symbols) for $12 \times 12$ superlattices at $T=0.10$ as a function of $U$.
Dashed  line is  perturbation theory analytical result,  from Ref. \onlinecite{White89b}.
}
\label{fig:txU0vsU} 
\end{figure}

We can gain further insight by analyzing the effective hopping separately within  free and repulsive layers. This is done in Fig. \ref{fig:txU0vsU}, where  $t^U_{x,eff}/t_x$(filled symbols)  and $t^0_{x,eff}/t_x$ (empty symbols)  are  normalized  by the number of repulsive and free sites, respectively.  The effective hopping along the repulsive sites is independent of the superlattice pattern for small $U$  and  follows closely the analytical results for perturbation theory in the homogeneous system \cite{White89b}. As $U$ is increased there are small deviations from the homogeneous counterpart. From the similarity  of the effective hopping in the homogeneous system and along the repulsive layers, we can infer that transport properties are essentially the same in both, leading to an insulating behavior of the fermions along the correlated layer.  Therefore, the main contribution to the transport in the direction parallel to the layers comes from the non-interacting sites, where $t^0_{x,eff}/t_x>1$. Fig. \ref{fig:txU0vsU} shows that  as $U$ is increased the contribution to the kinetic energy due to hopping between free sites is enhanced. When comparing  different  patterns we can see that the enhancement in $t^0_{x,eff}$ is optimized in lattices with narrow non-interacting layers ($L_0=1$) separated by  wide repulsive ones. The picture emerging from Fig. \ref{fig:txU0vsU} is that, while the repulsive layers are insulating, transport along the non-interacting layers is actually improved by the presence of the neighboring repulsive sites and increasing interaction strength among them.

The effective hopping cannot be used as an order parameter for the metal-insulator transition as it only approaches zero for $U \to \infty$ in the homogeneous case, whereas the system is known to be in a Mott insulating state for any non-zero value of $U$. Among the available quantities to characterize the transport properties of the HM within DQMC simulations,  the Drude weight is the one  less affected by finite size effects \cite{Mondaini12}.  It can be determined by \cite{Scalapino93}
\begin{equation}
\frac{D_\alpha}{\pi e^2}\equiv \lim_{T,m\to 0} [\langle-k_\alpha\rangle-\Lambda_{\alpha\alpha}(\textbf{q}=0,i\omega_m)],
\label{eq:Drude}
\end{equation}
where $\langle k_\alpha\rangle$ is the contribution to the average kinetic energy from fermions hopping along the  $\alpha$ direction and $\Lambda_{\alpha\alpha}(\textbf{q}=0,i\omega_m)$ is the long wavelength limit of the current-current correlation function with $\omega_m=2m\pi T$ the Matsubara frequency and $\alpha=x$ or $y$.
In principle the Drude weight as defined by (\ref{eq:Drude}) only provides information about the ground state of the system. Here we calculate  temperature dependent approximants to $D$ and analyze how they converge as the temperature is lowered.

\begin{figure}[t]
\vspace{-1.0cm}
{\centering\resizebox*{3.5in}{!}{\includegraphics*{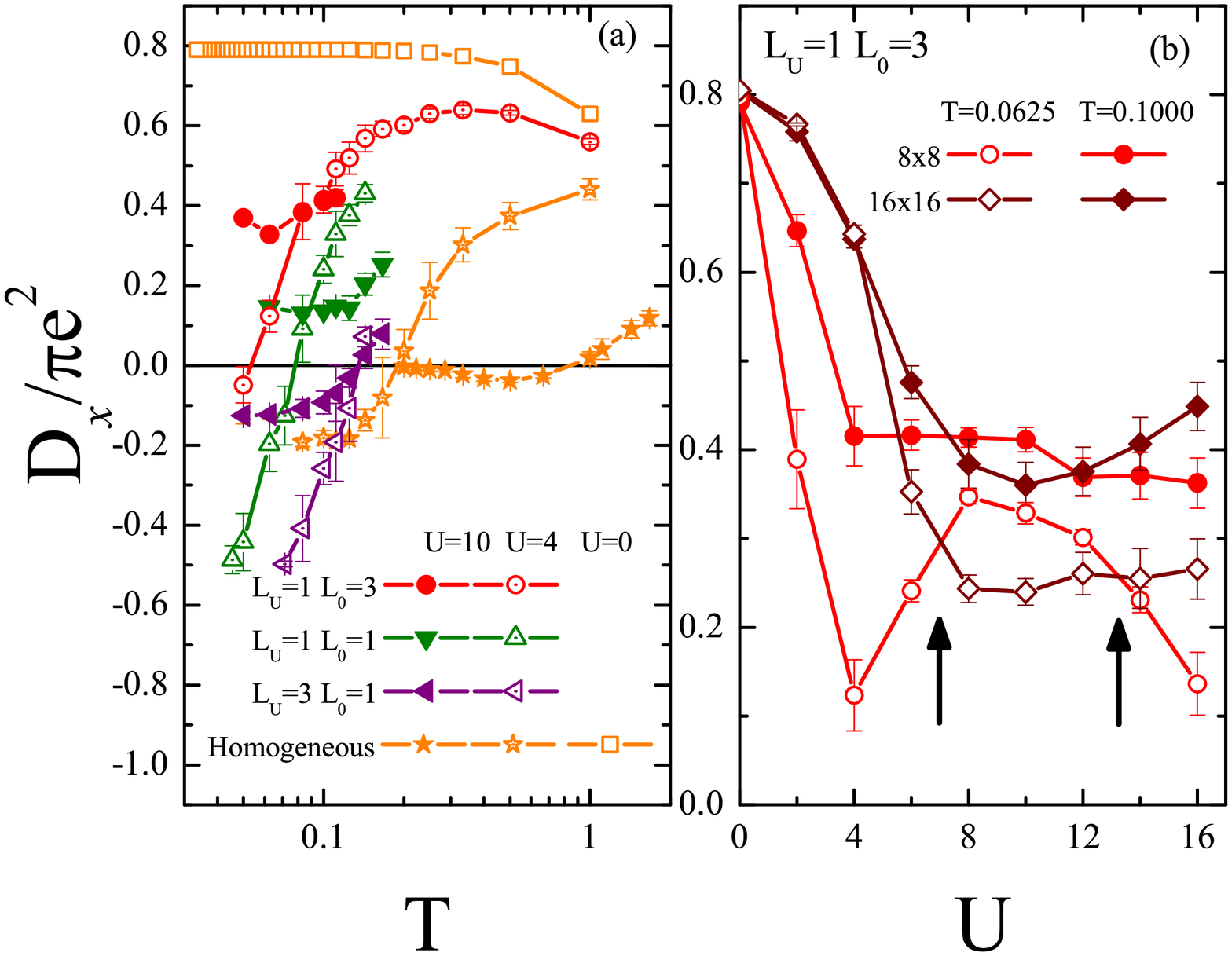}}}
\vspace{-0.7cm}
\caption{(Color online)
(a) Temperature dependence of the Drude weight along the direction of the layers, for $8\times 8$ lattices for different SL's with $U=4$ and  $U=10$  and homogeneous system with $U=0$, 4 and  10; (b) $U$ dependence of the Drde weight for the $L_U=1$ $L_0=1$ superlattice at $T=0.0625$ and $T=0.1000$.
}
\label{fig:Drude} 
\end{figure}

The Drude weight takes different values along the two preferential directions. Across the layers it displays an insulating character for all superlattices considered (not shown) in agreement with the effective hopping behavior. In Fig. \ref{fig:Drude}(a) we plot the Drude weight  along the direction of the layers for a fixed system size ($8\times 8$)  for different patterns as wells as for the homogeneous system.  The Drude weight for the $U \ne 0$ homogeneous system (stars) shows the expected insulating behavior; as the temperature is reduced $D/\pi e^2 \to 0$ at a given temperature, we shall call $T^*$. For $U=4$ (dot centered stars),  $T \sim 0.2$ and for $U=10$ (closed stars),  $T^* \sim 0.85$. $T^*$ is related  to the charge gap and increases as $U$ increases.
Our values of $T^*$ are in agreement with the results of Ref.\ \onlinecite{Vekic93} where this temperature is obtained in a completely different fashion, by the gap emergence in the spectral weight function.
The  negative values of $D$ are  a feature of the finite system size\cite{Scalapino93},  and converge to zero  in the thermodynamic limit  \cite{Mondaini12}. For the non-interacting homogeneous system (open squares) $D/\pi e^2=0.79$ as $T \to 0$, indicating an isotropic metallic behavior.

Let us now turn to the superlattices, with an intermediate interaction strength $U=4$ (dot centered symbols), the Drude weight goes to zero, signaling an insulating ground state; $T^*$ depends on the interaction pattern and increases as $L_U/L_0$ increases, being smaller than its homogeneous counterpart.  Contrary to what happens in the homogeneous system, increasing $U$ does not increase the charge gap, see closed symbols in Fig. \ref{fig:Drude}(a). For the system with $L_U=3$ and $L_0=1$,  $T^*$ is roughly the same for $U=4$ and $U=10$, whereas for the systems with $L_U/L_0 \le 1$ the Drude weight  stabilizes at non-zero values for the lower temperatures achieved,  and approaches that of the non-interacting case as $L_U/L_0$ is reduced.

In Fig. \ref{fig:Drude}(b) we study size and temperature effects on the Drude weight for the $L_U=1$ $L_0=3$ system. As $D_x/ \pi e^2$ evolves with $U$ we can divide the system's behavior in three regions, determined by the crossings  between  $8 \times 8$ and $16 \times 16$ data.  Let us call the values of $U$ where the crossings take place $U_{C1}$ and $U_{C2}$. For  $U < U_{C1} $ one can see that  $D_x/ \pi e^2$ is larger for the larger system size,  pointing  to a non-zero contribution as $T \to 0$ and $L \to \infty$, the characteristic metallic behavior.
 For $U_{C1} < U < U_{C2}$ we have the opposite behavior; as the system size is increased, the Drude weight decreases, leading to a vanishing contribution in the thermodynamic limit. Note that this insulating region increases as the temperature decreases. As $U$ increases even further and the region beyond $U_{C2}$ is reached, the behavior is similar to that of the first region: $D_x/ \pi e^2$ is larger for the larger system size, and the system is once again metallic.
The values of  $U_{C1}$ and $U_{C2}$ depend on temperature and are a likely to change when larger system sizes are considered. Nonetheless, the existence of two critical $U$ values leading to a  metal to  Mott insulator transition for small $U$, followed by a transition from an insulating to a metallic state along the direction of the layers at large $U$ is supported not only by the Drude weight data, but also from the effective hopping and double occupation.

In summary, we have employed Determinantal Quantum Monte Carlo simulations to study the effect of spatially varying interactions in two dimensional half-filled systems strongly interacting fermions.  We have found that, as expected, when interactions are turned on the system undergoes a metal to Mott insulator transition. Although the system is particle-hole symmetric and charge is evenly distributed among free and repulsive sites, the double occupation is not constant throughout the lattice. Double occupation on non-interacting sites  are influenced by the neighboring repulsive sites, first decreasing as $U$ increases and then increasing to return to the uncorrelated value as $U \to \infty$. This upturn  in the  double occupation  signals the decoupling between free and repulsive layers that takes place at large values of $U$. Evidence from double occupation, effective hopping and Drude weight show that, when this decoupling occurs, the system undergoes an exotic transition from a Mott insulator to metal, driven by fermionic correlations. We hope this results will encourage the experimental realization of  such spatially varying interactions in optical lattices.

\begin{acknowledgments}
We are grateful to R. R. dos Santos  and R. T. Scalettar for useful discussions.
Financial support from the Brazilian agencies CNPq, CAPES, FAPERJ, and INCT on Quantum Information is  gratefully acknowledged.
\end{acknowledgments}

\end{document}